\documentstyle[psfig]{mn}
\def\ref{\par\noindent\hang}

\def\spose#1{\hbox to 0pt{#1\hss}}
\def\approxlt{\mathrel{\spose{\lower 3pt\hbox{$\sim$}}
	\raise 2.0pt\hbox{$$<$$}}}
\def\approxgt{\mathrel{\spose{\lower 3pt\hbox{$\sim$}}
	\raise 2.0pt\hbox{$>$}}}

\def\multleft#1{\hbox to size{\vbox {\halign {\lft{##}\cr #1}}\hfill}\par}
\def\multright#1{\hbox to size{\vbox {\halign {\rt{##}\cr #1}}\hfill}\par}
\def\Mdot{\hbox{$\dot M$}}

\def\today{\ifcase\month\or January\or February\or March\or April\or May\or
      June\or July\or August\or September\or October\or November\or December\fi
      \space\number\day, \number\year}
\def\$<${\thinspace}
\def\s{\hbox{\phantom{5}}}	

\def\boxit#1{\vbox{\hrule\hbox{\vrule\kern3pt\vbox{\kern3pt
          #1 \kern3pt}\kern3pt\vrule}\hrule}}

\def\cm{{\rm\thinspace cm}}

\def\erg{{\rm\thinspace erg}}

\def\keV{{\rm\thinspace keV}}
\def\km{{\rm\thinspace km}}

\def\Msun{\hbox{$\rm\thinspace M_{\odot}$}}
\def\pc{{\rm\thinspace pc}}

\def\s{{\rm\thinspace s}}
\def\yr{{\rm\thinspace yr}}

\def\ergpcmsqps{\hbox{$\erg\cm^{-2}\s^{-1}\,$}}

\def\ergps{\hbox{$\erg\s^{-1}\,$}}

\def\kmps{\hbox{$\km\s^{-1}\,$}}

\def\pcmsq{\hbox{$\cm^{-2}\,$}}

\def\psqcm{\hbox{$\cm^{-2}\,$}}

\title[XRB and Nuclear Starbursts]
{Do nuclear starbursts obscure the X-ray Background?}
\author[A.C. Fabian et al]
{\parbox[]{6.5in}{A.C. Fabian$^1$, X. Barcons$^{1,2}$, O. Almaini$^1$
and K. Iwasawa$^1$}\\
\\
$^1$ Institute of Astronomy, Madingley Road, Cambridge CB3 0HA\\
$^2$ Instituto de F\'{\i}sica de Cantabria (Consejo Superior de
Investigaciones Cient\'{\i}ficas - Universidad de Cantabria), 39005
Santander, Spain} 

\date{25 March 1998}

\begin{document}

\maketitle

\begin{abstract}
We propose a model for the source of the X-ray background (XRB) in
which low luminosity active nuclei ($L\sim 10^{43}\ergps$) are
obscured ($N\sim 10^{23}\, {\rm cm}^{-2}$) by nuclear starbursts
within the inner $\sim 100$~pc. The obscuring material covers most of
the sky as seen from the central source, rather than being distributed
in a toroidal structure, and hardens the averaged X-ray spectrum by
photoelectric absorption. The gas is turbulent with velocity
dispersion $\sim {\rm few}\times 100\, \kmps$ and cloud-cloud
collisions lead to copious star formation.  Although supernovae tend
to produce outflows, most of the gas is trapped in the gravity field
of the starforming cluster itself and the central black hole.  A hot
($T\sim 10^6-10^7$~K) virialised phase of this gas, comprising a few
per cent of the total obscuring material, feeds the central engine of
$\sim 10^7\, \Msun$ through Bondi accretion, at a sub-Eddington rate
appropriate for the luminosity of these objects. If starburst-obscured
objects give rise to the residual XRB, then only 10 per cent of the
accretion in active galaxies occurs close to the Eddington limit in
unabsorbed objects.

\end{abstract}

\section{INTRODUCTION}

The flat spectrum of the X-ray background (XRB) above 1~keV (Marshall
et al 1980; Gendreau et al 1995; Chen, Fabian \& Gendreau 1997) is not
simply accounted for by the integration of known classes of source,
which generally have much steeper spectra. A currently popular model
invokes intrinsic absorption in active galaxies with which to flatten
the observed X-ray spectrum (Setti \& Woltjer 1989; Madau, Ghisellini
\& Fabian 1994; Celotti et al 1995; Comastri et al 1995). The
intrinsic absorption must range from column densities of about
$10^{22} - 10^{24}\psqcm$, and must cover most (at least $\sim
2/3$)\footnote{Since the XRB has a flat spectrum right down to 1~keV,
the fraction of sources with absorption exceeding $10^{21}\, \pcmsq$
must approach 90 per cent; see Section 3} of the sky
as seen by the source itself. The combination of different levels of
absorption and redshift of the objects can then lead to the observed
power-law background spectrum in the 1--10~keV band. The rollover in
the spectrum at about 30~keV is due redshift acting on an intrinsic
100~keV break, such as is seen in nearby active galaxies (Zdziarski et
al 1995).

The geometry of the obscuring material within a typical, X-ray
background-contributing, active galaxy is unclear. The nucleus must be
mostly surrounded by a typical column density of say
$10^{23}\psqcm$. If the material is freely orbiting the nucleus, then
it should soon collide with itself, dissipate, and flatten into a
disk, unless either a) the orbits are carefully arranged or b) there
is sufficient energy to continually throw matter into a wide range of
orbital inclinations. We note that what evidence there is for a torus
of molecular material around active galaxies does not indicate that
its high column density part covers much of the Sky. Indeed, models
where the obscuring material is extended on scales $\sim 100\, \pc$
(Granato et al 1997) account for the infrared properties of Seyfert
galaxies more successfully than ones which postulate a thick compact
($<1\, \pc$) torus (Pier \& Krolik, 1992, 1993).

The first case a) may be
accounted for by a warped disk. Pringle (1996) and Maloney, Begelman \&
Pringle (1996) have shown that the outer parts of irradiated accretion
disks are unstable to warping, the final result of which is that much of
the sky is covered by the warp. The details are currently uncertain as to
whether most of the sky can be covered and whether such high column
densities can be attained in the warped material. We do not pursue that
further here. The second case b) may be accounted for by a nuclear
starburst, the supernovae of which can provide the energy to push clouds
around and so obscure the nucleus. It is that possibility we explore
further here.

There is indeed much empirical evidence for a connection between
nuclear starbursts and active galactic nuclei (Terlevich \& Melnick
1985, Terlevich, D\'{\i}az \& Terlevich 1990; Perry \& Williams 1993,
Heckman et al 1995, 1997; Maiolino et al 1997). The extra featureless
ultraviolet continuum in Seyfert 2 galaxies (FC2) appears to be due to
a nuclear starburst (Terlevich \& Cid-Fernandes 1995; Heckman et al
1997). Turner et al (1997, 1998) in a discussion of the absorption
properties of Seyfert 2 galaxies propose that the starburst may be
involved there. The striking similarity between the evolution of the
luminosity density due to star formation (Madau et al 1996) and that
due to QSOs (Boyle \& Terlevich 1998, Dunlop 1998) also suggests a
close link between star formation processes and the fueling of QSO
accretion. In addition, some sort of starburst activity is inevitable
when an AGN forms, since vast amounts of material (e.g., triggered by
merging) are funneled into the core of the galaxy giving rise
ultimately to the central engine.  It is then likely that this
extended obscuring starburst is a common feature of all AGN during at
least some phase, although for the most massive and luminous objects
it might last only for a short time.
 
We make here the proposal that the formation of massive stars in the
inner $\sim 100$ pc of a central massive black hole is instrumental in
both fuelling the central engine by accretion and obscuring it by
distributing cold clouds all around that engine. Since most of the
X-ray accretion power in the Universe emerges in the XRB (see
e.g. Chen et al 1997) then it is in the obscured starburst mode that
most of it takes place.

An obscuring starburst also accounts for the optical, narrow-line,
appearance of a population of faint hard X-ray sources (NLXGs; Boyle et
al 1995; Roche et al 1995; Carballo et al 1995; Griffiths et al 1996;
McHardy et al 1998; Hasinger 1996; Almaini et al 1996; Romero-Colmenero
et al 1996) which dominate at faint flux levels and so may provide most
of the XRB (see, however, Hasinger et al 1998 and Schmidt et al 1998
who cast some doubts on the reality of these objects as a class
different to AGN). The broad-line Seyfert spectrum will not be detectable,
unless we have a favourable line of sight, or observe in the mid-infrared in
an object where the column density is not too high (Granato, Danese \&
Franceschini 1997). The combination of a
starburst spectrum and a Seyfert narrow-line spectrum will make
classification of the objects ambiguous (Iwasawa et al 1997b).

There are various examples of galaxies whose spectrum is dominated by
a starburst/LINER component at all wavelengths except at hard X-rays
where the AGN shows up, obscuration preventing its detection at lower
energies. Among them, NGC4945 is obscured by a column $\sim 5\times
10^{24}\, \pcmsq$ (Iwasawa et al 1993) and NGC 6240 by a column $>
{\rm few}\times 10^{24}\, \pcmsq$ (Iwasawa \& Comastri 1998).

\section{The distribution of the obscuring material}

We begin by assuming that the nuclear region of the galaxy has a
radius of $100R_2\pc$ and that it contains clouds with a total column
density of $10^{23}N_{23}\psqcm$ and a covering fraction approaching
unity. The circular orbital velocity is $10^2v_2\kmps$ and we now take
$R_2$ and $v_2$ $\sim 1$ to a few. We assume that this material sits
in an isothermal sphere, for which the gravitational effects of the
central black hole are negligible (see later).  Baryon dissipation
leads to a large baryon overdensity, and therefore dark matter
effects can be neglected. Fig.~1. shows that the
values assumed for the velocity and the total column density are
plausible, within an isothermal sphere approximation, with a
scale-length parameter $\alpha=100 R_2\pc$ and mass density $\lambda
(\Msun\, \pc^{-3})$. An approximation of the form $\rho(r)=\lambda
\left[ 1+(r/\alpha)^2/6\right]^{-1}$ has been used. The total mass of
the star cluster is likely to be $10^7-10^9\, \Msun$ (it has been
truncated at 3 scale radii).

\begin{figure} 
\centerline{\psfig{figure=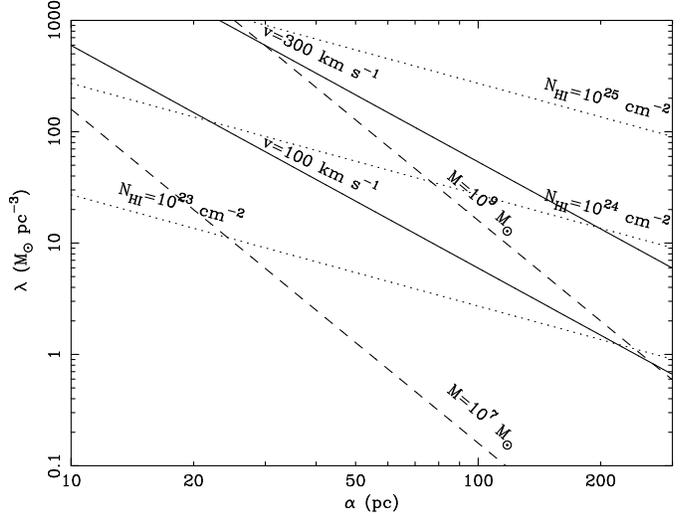,width=0.5\textwidth,angle=270}}
\caption{Various physical parameters for the obscuring material
(velocity, column density from the centre outwards, mass) are shown as
a function of the scale length and mass density in an isothermal
sphere truncated at $3\alpha$. The column density values shown include
all the mass in the star cluster, of which the absorbing gas will only
be a fraction.}
\end{figure}

The orbital kinetic energy of the clouds is $3\times 10^{54} N_{23} R_2^2
v_2^2\erg$; half the orbital time, on which collisions can be expected if
the clouds are on highly inclined orbits, is $3\times 10^6 R_2
v_2^{-1}\yr$. After this time most of the clouds will have collided and
formed a more flattened configuration. Thus a power of $\dot E\sim
3\times 10^{40} N_{23} R_2 v_2^3\ergps$ must be supplied after a few million yr
to keep the covering fraction high.

The obvious origins for such power are the winds from massive stars,
supernovae and the active nucleus. The problem with the latter is coupling
any of its power into inclining the orbital motions of clouds in a random
sense. The blast wave of supernovae is likely to lead to a highly radiative
shock in the dense environment considered here. However the random
positioning of supernovae and massive stars is likely to mean that at least
some of the power goes into turbulence within the region. This could have
the desired effect.

Note that we are dealing here with a starburst in the potential well
of the (stellar plus black hole) nucleus, which will tend to trap
material than simply lead to outflow, as for off-nucleus
starbursts. Outflows may and probably do occur (at least in favourable
directions where the column of gas is smaller, as in the superwinds
seen in local starburst galaxies, Heckman, Armus, Miley 1990) but more
energy is required. It would be difficult to eject all the surrounding
gas, since the more gas is ejected via supernovae, the less stars and
supernovae are formed.

One important effect of the combined gravity field created by the black
hole and the star cluster is to drive the distribution of material to
an axisymmetric distribution only in the innermost region where the
black hole dominates (S. Sigurdsson, private communication). Further
away, the distribution is expected to be triaxial and therefore the
obscuring material will tend to cover the central source in a rather
isotropic way.

Here we postulate that a few per cent of the energy of all supernovae
(say 1 per 100~yr) in the region goes into turbulence within the inner
$\sim 100$~pc. This can then randomize the orbits of the clouds there
and so cause a large covering fraction.

The frequent collisions between clouds is one trigger for star formation
within the clouds, thus the situation once begun may continue until most of
the gas is used up.

The starburst thus perpetuates itself, and obscures the nucleus. Of course
the collisions will lead to a multiphase medium present, some of which is
accreted by the central black hole. It is therefore a source of fuel for the
central engine. Actually, hot gas flows are recognised as a very
efficient way to fuel AGN and at the same time extract angular
momentum from the accretion disk (Shlosman, Begelman \& Frank 1990). 

Assuming a $10^7\, M_7\, {\rm M}_{\odot}$ black
hole and that a fraction $10^{-2}f_2$  of the gas is in virial
equilibrium (hot phase) with a temperature $T\approx 4\times 10^6\,
v_2^2\, {\rm K}$.  The density of the hot phase is therefore $\approx
0.5\, R_2^{-1} N_{23} f_2\, {\rm cm}^{-3}$.  Bondi accretion onto the
central mass, assuming the gas is isothermal, gives 
\[
\Mdot=1.5\times 10^{-3} M_7^2 N_{23} R_2^{-1}f_2v_2^{-3}\, {\rm
M}_{\odot}\, {\rm yr}^{-1}
\]
and for a typical 10 per cent efficiency this is converted into a
luminosity $\approx 8\times 10^{42}M_7^2 N_{23} R_2^{-1}f_2v_2^{-3}\,
{\rm erg}\, {\rm s}^{-1}$.  This is of the order of the X-ray
luminosity of a NLXG. 

The distribution of gas changes abruptly when the ratio of the black
hole mass to the mass distributed in a sphere of one scale-length
radius ($M (\alpha)$)
\[
{M_{BH}\over M (\alpha)}=0.12 R_2\, v_2^{-2}\, M_7
\]
approaches or exceeds unity: it goes from gravity being dominated by
the star cluster itself to being dominated by the black hole (Huntley
\& Saslaw 1975).  In the latter case, all the gas virtually sinks to
subparsec distances and nothing like an extended distribution of gas
and stars survives.  A velocity of a few $100 \kmps$ would be
difficult to maintain and such velocities are needed since the NLXGs
(which are our prototype objects) exhibit only narrow emission lines.
The transition between obscured starburst to `normal' broad-line AGN
is likely to happen when the mass of the central engine exceeds a few
$\times 10^8\, \Msun$.

Moreover, the ratio of Bondi accretion to the Eddington limit is
\[
{\Mdot\over\Mdot_{Edd}}=7\times 10^{-3}\, M_7\, N_{23}\, R_2^{-1}\,
f_2\, v_2^{-3}
\]
which increases with the mass of the central engine. When $M_7$
exceeds $\sim 10$, the gas sinks and $R_2$ significantly reduces,
in which case the accretion rate approaches or exceeds the Eddington
limit. The central engine will radiate close to the Eddington limit
and the remaining accreted material will be blown away from the
nucleus.  Again this is likely to lead to a normal unobscured
broad-line AGN. Note that the lifetime of the starburst is unlikely to exceed
the Salpeter time for doubling the black hole mass by Eddington-limited
accretion. We therefore ignore variations in the black hole mass.

\section{Discussion and Conclusions}

Nuclear starbursts can play a major role in obscuring low-luminosity
AGN.  The emerging X-ray spectrum will be moderately to highly
absorbed by the cold and warm gas in the star cluster, thus providing
the flat spectral shape that is needed to produce the XRB.  The column
of absorbing gas is likely to vary vith viewing direction in a
`random' way due to supernovae produced in the starburst.  The average
high fraction of the sky that the majority of the faint X-ray sources
have to see covered by absorbing gas arises in material related to the
star cluster rather than in a hypothetical torus, whose thickness to
radius ratio must be large. Arguments favouring a geometry for the
obscuring material more isotropic than toroidal have been put forward
before. Turner et al (1998)  find no obvious
dependence of the X-ray properties of NLXGs with the orientation of
the host galaxy. Iwasawa et al (1995) also suggested that the
obscuring material in the Seyfert 2 galaxy $IRAS$ 18325-5926, which is
obscured by a column in excess of $10^{22}\, \pcmsq$ but shows no
X-ray reflected component, is likely to be isotropically distributed
around the central engine.

Indeed the question remains, as in all obscured AGN models for the
XRB, as to why the distribution of obscuring material along different
lines of sight in the population of these objects is exactly tuned in
such a way that it gives rise to the featureless spectrum of the
XRB. There is no physical principle which leads to this and some
fine-tuning of this distribution is required to produce a detailed
model for the XRB.

The optical/UV properties of the highly absorbed AGN will be
dominated by the starburst itself.  Collisions between cold clouds,
occuring at several $\times 100 \kmps$ will lead to radiative shocks
with some emission at soft X-ray energies.  Therefore, the obscuring material
will also contribute to the observed soft X-ray emissivity, making
some of these objects visible in the ROSAT PSPC band in spite of being
absorbed.

Near infra-red spectroscopy of NLXGs also lends support to this hybrid
(i.e., accretion onto a black hole plus an obscuring starburst)
hypothesis. While a number of NLXGs show unambiguous evidence for
hidden AGN (eg. highly ionized coronal lines), the non-detection of
broad Paschen $\alpha$ requires obscuring columns with $N_H >
10^{23}$~cm$^{-2}$ (Almaini et al 1998). Such thick columns are
inconsistent with the large X-ray luminosities observed in ROSAT,
unless there is an additional source of soft X-ray flux. A
contribution from the activity in the obscuring starburst provides a
natural explanation.

Gas columns along typical lines of sight in these objects are modest
and for normal dust to gas ratios the absorbing material will be
optically thin in the mid- and far-infrared (Granato et al 1997) where
most of UV and soft X-ray reprocessed radiation will be emitted in a
rather isotropic fashion.  Surveys at these wavelengths (particularly
at $10-15\, \mu$m) should be especially efficient in finding these
objects. Barcons et al (1995) measured the average ratio of X-ray
(2-10~keV) to mid-infrared emission (the {\it IRAS} 12$\mu$m band) for
12$\mu$m-selected type 1 and type 2 AGN.  The starburst obscured AGN
are likely to have $f(5\, \keV)/f(12\, \mu{\rm m})\sim
2\times 10^{-7}$, flux ratio that was found for Seyfert 2 galaxies, and
therefore an obscured AGN with a 2-10~keV X-ray flux of $10^{-14}\,
{\rm erg}\, {\rm cm}^{-2}\, {\rm s}^{-1}$ is likely to emit at a level
of $\sim 3$~mJy at 10-15$\mu$m.  The {\it European Large Area Infrared
Space Observatory Survey} (ELAIS, Oliver et al 1997) is reaching a 
sensitivity of $\sim 2$~mJy at 15$\mu$m over a survey area of 13${\rm
deg}^2$ (some 1000-2000 objects are expected in total). Even if at a
2-10~keV flux of $10^{-14}\, \ergpcmsqps$ only a minor fraction of the
sources are starburst obscured AGN, they will certainly show up in
large numbers in the ELAIS survey.

The recently commissioned Sub-mm Common User Bolometer Array (SCUBA)
at the James Clerk Maxwell Telescope in Hawaii offers an ideal
opportunity to detect obscured AGN at high redshifts. The re-radiation
of nuclear energy in the far infra-red is expected to produce a
thermal emission peak at $\lambda \sim 60-100\mu$m, which is
redshifted into the sub-millimetre region. A starburst-obscured AGN at
redshift $z=2$ with a 2-10~keV X-ray flux of $5\times10^{-15}\,
\ergpcmsqps$ is expected to give a 350$\mu$m flux of $\sim
20$mJy. Deep SCUBA observations of hard X-ray selected fields
(eg. with ASCA, AXAF or XMM) should reveal these obscured objects.

As stated, we expect that the very massive black holes will appear as
quasars. Sometimes however a massive black hole ($M_{BH}>10^8\Msun$) may have
a more massive surrounding star cluster in which case our scenario may
apply and the active nucleus will be hidden. This may explain the low
level of X-ray emission from IRAS galaxies in general and the
hyperluminous galaxies (e.g. IRAS F15307+3252) in particular (Fabian et al
1996). The X-ray upper limit to that object requires less than $2\times
10^{-4}$ of its power to emerge as X-rays which means that any
quasar-like nucleus must be smothered from all directions.

\begin{figure} 
\centerline{\psfig{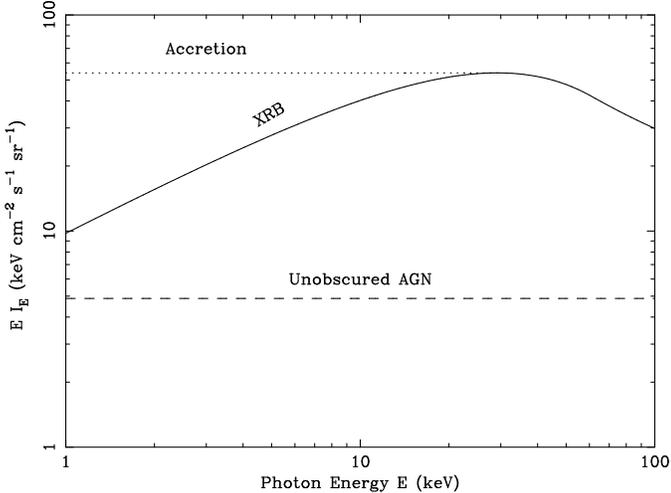}}
\caption{The energy content of the XRB per unit logarithmic energy
interval (solid line) as a function of photon energy.  The dashed
curve shows the contribution of unabsorbed AGN, where it is assumed
that they make 50 per cent of the XRB at 1~keV. There is probably a
roll-over in their contribution at energies $> 30-50\, \keV$ which is the
exponential cutoff observed in local AGN at $\sim 100\, \keV$ redshifted
out to $z\sim 2$. The dotted line shows the total energy produced by
all objects, before obscuration is taken into account.}
\end{figure}

A geometrically thin disk structure is expected in obscured AGN in our
model on scales of $<1\pc$ where angular momentum begins to
be important. Compton scattering, in some cases in an optically-thick
medium, is necessary to explain the X-ray emission of some Seyfert 2
galaxies (see Iwasawa et al 1997a for a detailed study of NGC 1068 and
Turner et al 1997 for Mrk 3 among other examples). In the case of NGC
1068 the pc-scale disk structure has been mapped at radio
wavelengths (Gallimore, Baum \& O'Dea 1997). Our model can therefore
account for both Compton-thin and Compton-thick Seyfert 2
galaxies. Much of the extended torus attributed to these objects is
just the starburst obscuration, but a planar Compton-thick inner part
lies along our line of sight in some of them.

If the model proposed here describes correctly the sources of the
residual XRB, it has then important implications on how accretion
occurs in the Universe.  Fig.~2 shows the distribution of the energy
in the XRB as a function of photon energy in the range 1 to 100 keV
(Fabian \& Barcons 1992). Unobscured AGN (QSOs and Seyfert 1s) make
$\sim 50$ per cent of the XRB at 1~keV (Hasinger 1996, McHardy et al
1998, Hasinger et al 1998, Schmidt et al 1998), with an energy
spectral index $\sim 1$. If the `shoulder' in the XRB (Fig.~2)  is due to
obscuration (mostly photoelectric absorption of the photons with
energies $< 30\, \keV$), the energy content at 30 keV provides a measure
of the total energy produced by accretion in AGN, for the same
underlying power law spectrum.  What is remarkable then, is that only
10 per cent of the accretion occurs in unobscured objects (assuming
that before it is absorbed by the surrounding material, the spectrum
generated by accretion in both cases is similar). The remaining 90 per
cent occurs at sub-Eddington accretion rates, and a relevant fraction
is absorbed and re-radiated at longer wavelengths, particularly in the
infrared. This means that most of the sky as seen from an average
active nucleus is obscured.

Moreover, given the requirement that these low luminosity
AGN are at least one order of magnitude fainter than typical broad line
objects (see Section 2), we predict that starburst obscured AGN may
outnumber brighter, unobscured AGN by a factor of 100 or more if they are
to account for the remainder of the XRB. This estimate is in good
agreement with the number count predictions for QSOs and NLXGs based on
extrapolating deep ROSAT observations (Almaini \& Fabian 1997). 

If most of accretion in the Universe is highly obscured, then the
amount of emitted power per unit galaxy based on optical or UV QSO
luminosity functions (So{\l}tan 1982, Phinney 1997), and therefore the
mass in black holes in AGN, might have been underestimated. This is
due to the fact that 90 per cent of the accretion power would be obscured
and re-radiated in the infrared.  Also, if most AGN are Advection
Dominated Accretion Flows (ADAFs), as it has been proposed by Di
Matteo \& Fabian (1997) as the source of the XRB, the implied low mass
to energy conversion efficiency also means that the black hole masses
would have to be larger, in agreement with the local estimates.

\section*{acknowledgments}

We thank Steinn Sigurdsson for very interesting comments.
ACF thanks the Royal Society for support.  XB acknowledges
partial financial support provided by the DGES under project PB95-0122
and funding for his sabbatical at Cambridge under DGES grant PR95-490.

\end{document}